\RequirePackage{fix-cm}
\documentclass[10pt]{article}         
\usepackage{amsmath}\usepackage{amsmath,amsfonts,amssymb,amsthm,url,array}
\usepackage{mathbbol}
\usepackage{abstract}
\usepackage{hyperref}
\usepackage[utf8]{inputenc}
\usepackage[T1]{fontenc}

\usepackage[sort]{natbib}

\usepackage{setspace}
\usepackage{mathptmx}
\usepackage{titlesec}
\titleformat{\paragraph}
{\normalfont\normalsize\bfseries}{\theparagraph}{1em}{}
\titlespacing*{\paragraph}
{0pt}{3.25ex plus 1ex minus .2ex}{1.5ex plus .2ex}
\usepackage{optidef}
\usepackage{mathrsfs} 
\usepackage{graphicx}
\allowdisplaybreaks
\begin{document}

\title{A Proposal for Multi-asset Generalised Variance Swaps}

\author{
    Subhojit Biswas\\
    \textit{Indian Statistical Institute, Kolkata}\\
    \textit{subhojit1016kgp@gmail.com}\\
    \\
    \\
    Diganta Mukherjee\\
    \textit{Sampling and Official Statistics Unit}\\
    \textit{Indian Statistical Institute, Kolkata}\\
    \textit{digantam@hotmail.com}\\
    \thanks{We thank Indranil Sengupta for helpful comments and suggestions which has helped improve the exposition considerably. The usual caveat applies. }
}

\maketitle

\begin{abstract}
This paper proposes swaps on two important new measures of generalized variance,  namely the maximum eigenvalue and trace of the covariance matrix of the assets involved. We price these generalized variance swaps for financial markets with Markov-modulated volatilities. We consider multiple assets in the portfolio for  theoretical purpose and demonstrate our approach with numerical examples taking three stocks in the portfolio. The results obtained in this paper have important implications for the commodity sector where such swaps would be useful for hedging risk. 
\end{abstract} 



\section{Introduction}

Covariance and correlation swaps are among recent financial products which are useful for volatility hedging and speculation using two different financial underlying assets. For example, option dependent on exchange rate movements, such as those paying in a currency different from the underlying currency, have an exposure to movements of the correlation between the asset and the exchange rate, this risk may be eliminated by using a covariance swap.

The literature devoted to the volatility derivatives is growing. The Non-Gaussian Ornstein-Uhlenbeck stochastic volatility model was used by \citeauthor{benth} (\citeyear{benth}) to study volatility and variance swaps. \citeauthor{broadie1} (\citeyear{broadie1}) evaluated price and hedging strategy for volatility
derivatives in the Heston square root stochastic volatility model and in \citeauthor{broadie2} (\citeyear{broadie2}) they compare result from various model in order to investigate the effect of jumps and discrete sampling on variance and volatility swaps. Pure jump process with independent increments return models were used by \citeauthor{carr1}(\citeyear{carr1}) to price derivatives written on realized variance, and
subsequent development by \citeauthor{carr1}(\citeyear{carr1}). This paper also provides a good survey on volatility derivatives. \citeauthor{fonseca}(\citeyear{fonseca}) analyzed the influence of variance and covariance swap in a market by solving a portfolio optimization problem in a market with risky assets and volatility derivatives. Correlation swap price has been investigated by \citeauthor{bossu1} (\citeyear{bossu1}) and \citeauthor{bossu2} (\citeyear{bossu2}) for component of an equity index using statistical method.

By definition, all the above methods can only consider a combination of two assets at a time. But in today's complex financial transactions, there is no reason why volatility of three or more assets will not be considered for contracting together. Thus, in this paper, we extend these methods to a situation where some generalized variance of a portfolio of assets can be contracted on. Taking cue from multivariate analysis, we look at two important measures of generalized variance, namely the {\it maximum eigenvalue} and {\it trace} of the covariance matrix of the assets involved.  
The objective is to price generalized variance swaps for financial markets with Markov-modulated volatilities. As an example, we consider stochastic volatility driven by a finite state continuous time Markov chain. To the best of our knowledge, this is the first attempt in extending the covariance swaps to a multidimensional situation.

We outline the problem and the theoretical results is section 2. First we look at case of the trace swap and in a subsequent subsection we discuss the eigenvalue swap with a target return constraint. The numerical examples are presented with real data in section 3. Finally section 4 concludes.

\section{Problem formulation}

Let us consider a financial market with two types of securities, the risk free bond and the stock. Suppose that the stock prices $(S_t)_{t \in R^+}$ satisfy the following stochastic differential equation
\begin{equation}
  d S_t = S_t(\mu dt + \sigma(x_t) dw_t) \nonumber \end{equation}
where $w$ is a standard Wiener process independent of the Markov process $(x_t)_t$. 

A portfolio consists of n stocks with the corresponding returns given by $\frac{dS_1}{S_1}$, $\frac{dS_2}{S_2}$,..... $\frac{dS_n}{S_n}$. The vector of individual returns has variances and co-variances involved with it. Let the portfolio return covariance matrix be given by
\[ \Omega = 
\begin{bmatrix}
    Cov (r_1,r_1) & Cov (r_1,r_2) & Cov (r_1,r_3) &.....& Cov (r_1,r_n) \\
    Cov (r_2,r_1) & Cov (r_2,r_2) & Cov (r_2,r_3) &.....& Cov (r_2,r_n) \\.......\\ Cov (r_n,r_1) & Cov (r_n,r_2) & Cov (r_n,r_3) &.....& Cov (r_n,r_n) 
\end{bmatrix}
\]
\[ \Omega = 
\begin{bmatrix}
    \sigma_{1}^2(x_t) & \rho_{(12)}\sigma_{1}(x_t)\sigma_{2}(x_t) & \rho_{(13)}\sigma_{1}(x_t)\sigma_{3}(x_t) &.....& \rho_{(1n)}\sigma_{1}(x_t)\sigma_{n}(x_t) \\
    \rho_{(21)}\sigma_{2}(x_t)\sigma_{1}(x_t) & \sigma_{2}^2(x_t) & \rho_{(23)}\sigma_{2}(x_t)\sigma_{3}(x_t) &.....& \rho_{(2n)}\sigma_{2}(x_t)\sigma_{n}(x_t) \\.......\\ \rho_{(n1)}\sigma_{n}(x_t)\sigma_{1}(x_t) & \rho_{(n2)}\sigma_{n}(x_t)\sigma_{2}(x_t)& \rho_{(n3)}\sigma_{n}(x_t)\sigma_{3}(x_t) &.....& \sigma_{n}^2(x_t) 
\end{bmatrix}
\]
\\
Let ($\Omega_S$, $\mathscr{F}$, $(\mathscr{F}_{t})_{t\in R^+}$, P) be a filtered probability space, with a right-continuous filtration $(\mathscr{F}_{t})_{t\in R^+}$ and probability P. The following two results allow us to associate $(x_t)_{t\in R^+},$ which is a Markov Process with generator Q, to a martingale and to obtain its quadratic variation \citeauthor{swischuk2} (\citeyear{swischuk2}). We refer to \citeauthor{swischuk1} (\citeyear{swischuk1}) for the proofs.\newline

\textbf{Proposition 1. }(\citeauthor{swischuk1},\citeyear{swischuk1}) Let $(x_t)_{t\in R^+}$ be a Markov process with generator Q and $f \in Domain (Q)$, then
\begin{equation}
    m_t^f = f(x_t) - f (x_0) - \int_{0}^{t} Q f(x_s) ds \nonumber
\end{equation}is a zero mean martingale with respect to the $\mathscr{F}_t: \sigma \{y (s) ; 0 \leq s \leq t\} $.\newline 

\textbf{Proposition 2. }(\citeauthor{swischuk1}, \citeyear{swischuk1}) Let $(x_t)_{t\in R^+}$ be a Markov process with generator Q, $f \in Domain (Q)$ and $(m_t^f)_{t\in R^+}$ its associated martingale, then
\begin{equation}
    <m^f>_t : = \int_{0}^{t} [Q f^2(x_s) - 2f(x_s) Q f(x_s)]ds \nonumber
\end{equation}
is the quadratic variation of $m^f$.\newline

\textbf{Proposition 3. }(\citeauthor{swischuk2}, \citeyear{swischuk2}) Let $(x_t)_{t\in R^+}$ be a Markov process with generator Q, $f, g \in Domain (Q)$ such that $fg \in Domain (Q)$. Denote by $(m_t^f)_{t\in R^+}$; $(m_t^g)_{t\in R^+}$ their associated martingale.Then
then
\begin{equation}
    <f(x.), g(x.)>_t : = \int_{0}^{t} [Q f(x_s)g(x_s) - f(x_s) Q g(X_s) -  g(x_s) Q f(x_s)]ds \nonumber
\end{equation}
is the quadratic variation of f and g.\newline

In our model the volatility is stochastic. Then it is interesting to study the property of $\sigma$ and in particular how to price derivative contracts on realized variance.
We consider $\sigma$ as a martingale as we are assuming that, regardless of a stock's current and past volatility, his expected volatility at any time in the future is the same as his current volatility.\newline

\textbf{Proposition 4. }(\citeauthor{swischuk2}, \citeyear{swischuk2}) Suppose that $\sigma \in Domain (Q)$. Then,

\begin{equation}
    E\{\sigma^2(x_t)|\mathscr{F}_u\} = \sigma^2(x_u) + \int_{0}^{t} E\{\sigma^2(x_s)|\mathscr{F}_u\} ds \nonumber
\end{equation}
for all $0\leq u \leq t$. The value of conditional expectation is given by
\begin{equation}
    E \{\sigma^2(x_t)|\mathscr{F}_u\} = e^{(t-u)Q}\sigma^2(x_u)   \nonumber
\end{equation}
If we remove the conditional part then,
\begin{equation}
    E \{\sigma^2(x_t)\} = e^{tQ}\sigma^2(x)   \nonumber
\end{equation}
where we have denoted $x_0 : = x.$

For the covariance terms we can write similar value for the expectation 
if we remove the conditional part. Then,
\begin{equation}
    E \{\sigma_1(x_t)\sigma_2(x_t)\} = e^{tQ}\sigma_1(x)\sigma_2(x).   \nonumber
\end{equation}

These results help us to derive the probability distribution of the eigenvalue that we discuss subsequently. We first look at the derivation of the trace swap.

\subsection{Swap using the trace of the covariance matrix}

As the first proposal, we consider the investor using the trace of the covariance matrix to develop the swap. The trace is given by 
\begin{equation}
    tr \hspace{2pt}\Omega (x_t) =  \sigma_{1}^2(x_t) +  \sigma_{2}^2(x_t) +  \sigma_{3}^2(x_t) + .... \sigma_{n}^2(x_t). \nonumber
\end{equation}

Now the price of the swap on trace is the expected present value of the payoff in the risk neutral world for the  assets we have considered

\begin{equation} 
    P_{trace} (x) = E \{e^{-rT}(tr \hspace{2pt}\Omega (x_t) - K_{\text{strike price}})\} \nonumber
\end{equation}
\begin{equation} 
    P_{trace}(x) = e^{-rT} E \{(tr \hspace{2pt}\Omega (x_t) - K_{\text{strike price}})\} \nonumber
\end{equation}

\textbf{Example: } Let us consider 3 stocks in the portfolio. Then the covariance matrix becomes

\[ \Omega = 
\begin{bmatrix}
    \sigma_{1}^2(x_t) & \rho_{(12)}\sigma_{1}(x_t)\sigma_{2}(x_t) & \rho_{(13)}\sigma_{1}(x_t)\sigma_{3}(x_t)\\
    \rho_{(21)}\sigma_{2}(x_t)\sigma_{1}(x_t) & \sigma_{2}^2(x_t) & \rho_{(23)}\sigma_{2}(x_t)\sigma_{3}(x_t)\\
    \rho_{(31)}\sigma_{3}(x_t)\sigma_{1}(x_t) & \rho_{(32)}\sigma_{3}(x_t)\sigma_{2}(x_t) & \sigma_{3}^2(x_t)
\end{bmatrix}
\]

\begin{equation}
    tr \hspace{2pt}\Omega (x_t) =  \sigma_{1}^2(x_t) +  \sigma_{2}^2(x_t) +  \sigma_{3}^2(x_t) \nonumber
\end{equation}

\begin{equation} 
    P_{trace} (x) = E \{e^{-rT}(tr \hspace{2pt}\Omega (x_t) - K_{\text{strike price}})\} \nonumber
\end{equation}
\begin{equation} 
    P_{trace}(x) = e^{-rT} E \{(tr \hspace{2pt}\Omega (x_t) - K_{\text{strike price}})\} \nonumber
\end{equation}
For our Markov modulated market, this becomes
\begin{equation} 
    P_{trace}(x) = \big(e^{-rT}\{\frac{1}{T}\int_{0}^{T}(e^{tQ} \sigma_{1}^2(x)) dt\}\big) + \big(e^{-rT}\{\frac{1}{T}\int_{0}^{T}(e^{tQ} \sigma_{2}^2(x)) dt\}\big) + \big(e^{-rT}\{\frac{1}{T}\int_{0}^{T}(e^{tQ} \sigma_{3}^2(x)) dt\}\big) - e^{-rT} K_{\text{strike price}} \nonumber
\end{equation}

\subsection{Swap using the largest eigenvalue}

The objective here is to define and derive the price of an eigenvalue swap. But we do not address the problem without an efficiency consideration as combinations of underlying assets for unconstrained variance may not be interesting as an investment destination.
So, here we assume that the investor considers the maximum eigenvalue of the covariance matrix, for a given expected mean return. For which we have to find the distribution.

Let the weights associated with the given stocks be
\[ \textbf{w(t)} = 
\begin{bmatrix}
    w_1(t) \\
    w_2(t) \\ w_3(t) \\ ....\\ w_n(t)
\end{bmatrix}
\]

The optimization problem can be written in the following structure

\begin{maxi}|l|
	  {\textbf{w(t)}}{\textbf{w(t)}^{T}\Omega \textbf{w(t)}}{}{}
	  \addConstraint{\textbf{w(t)}^{T}\textbf{w(t)}}{=1}{}
	  \addConstraint{\textbf{I}^T \textbf{w(t)}}{= 1}{}
	  \addConstraint{\text{E}(\textbf{R})^T\textbf{w(t)}}{= k,}{}\nonumber
\end{maxi}
\textbf{R} is the vector containing the expected return of the stocks. Overall the constraint can be combined as 
\begin{maxi}|l|
	  {\textbf{w(t)}}{\textbf{w(t)}^{T}\Omega \textbf{w(t)}}{}{}
	  \addConstraint{\textbf{w(t)}^{T}\textbf{I} \textbf{w(t)}}{=1}{}
	  \addConstraint{\textbf{A}^T \textbf{w(t)}}{= \bf{b},}{}\nonumber
\end{maxi}
where 
\[ \textbf{A} = 
\begin{bmatrix}
    \text{E}(\textbf{R}) & \textbf{I} 
 \end{bmatrix}
=
\begin{bmatrix}
    \mu_1 &  1 \\
    \mu_2 & 1\\
    \mu_3 & 1 \\
    .... & ..\\
     \mu_n & 1
\end{bmatrix}
\]

\[ \bf{b} = 
\begin{bmatrix}
   k \\
    1 
\end{bmatrix}.
\]

Here
$\Omega$ is a $n\times n$ and \textbf{A} is a $n\times 2$. We are going to simplify the first constraint and we are going to do a QR decomposition of the matrix \textbf{A}.\citeauthor{Ref3} (\citeyear{Ref3})
\[\textbf{P}^{T} \textbf{A} =
\begin{bmatrix}
     \textbf{R} \\ 0
\end{bmatrix},
\]

where \textbf{P} denotes an orthogonal matrix, and \textbf{R} is a upper triangular matrix
\[\textbf{A}^{T}\textbf{P} =
\begin{bmatrix}
     \textbf{R}^{T} & 0
\end{bmatrix}
\]

Multiplying both the sides by $\textbf{P}^{T}$
\[\textbf{A}^{T}\textbf{P}\textbf{P}^{T} =
\begin{bmatrix}
     \textbf{R}^{T} & 0
\end{bmatrix} \textbf{P}^{T}
\]
\[\text{or } \textbf{A}^{T} =
\begin{bmatrix}
     \textbf{R} & 0
\end{bmatrix} \textbf{P}^{T}
\]
The optimization problem now becomes,
\begin{maxi}|l|
	  {\textbf{w(t)}}{\textbf{w(t)}^{T}\textbf{P}\textbf{P}^{T}\Omega\textbf{P}\textbf{P}^{T} \textbf{w(t)}}{}{}
	  \addConstraint{\begin{bmatrix}
     \textbf{R}^{T} & 0 \end{bmatrix} \textbf{P}^{T}
 \textbf{w(t)}}{=\textbf{b}}
	  \addConstraint{\textbf{w(t)}^T \textbf{P}\textbf{P}^{T}\textbf{I}\textbf{P}\textbf{P}^{T} \textbf{w(t)}}{= 1.}{} \nonumber
\end{maxi}

We can now use the following definitions
\[\textbf{P}^{T}\Omega\textbf{P} =
\begin{bmatrix}
     \textbf{B} & \Gamma^{T} \\
     \Gamma & \textbf{C}
\end{bmatrix}
\]

where the dimensions of the matrix \textbf{B}, matrix $\Gamma$, matrix \textbf{C} will have the dimensions accordingly

\[\textbf{P}^{T}\textbf{w(t)} =
\begin{bmatrix}
     \textbf{q} \\
     \textbf{r}
\end{bmatrix}.
\]
Similarly the dimensions of the matrix \textbf{q}  and matrix \textbf{r} will be decided  accordingly

\[\textbf{w(t)}^{T}\textbf{P}\textbf{P}^{T} =
\begin{bmatrix}
     \textbf{q}^{T} & \textbf{r}^{T}
\end{bmatrix} \textbf{P}^{T}
\]
\[\textbf{w(t)}^{T} =
\begin{bmatrix}
     \textbf{q}^{T} & \textbf{r}^{T}
\end{bmatrix} \textbf{P}^{T}.
\]

Also, $\textbf{C} = \textbf{C} ^{T},$ so

\[\textbf{w(t)}^{T}\Omega\textbf{w(t)} = \textbf{w(t)}^{T}\textbf{P}\textbf{P}^{T}\Omega\textbf{P}\textbf{P}^{T} \textbf{w(t)} =
\begin{bmatrix}
     \textbf{q}^{T} & \textbf{r}^{T}
\end{bmatrix}
\begin{bmatrix}
     \textbf{B} & \Gamma^{T} \\
     \Gamma & \textbf{C}
\end{bmatrix}
\begin{bmatrix}
     \textbf{q} \\
     \textbf{r}
\end{bmatrix}
= \begin{bmatrix}
     \textbf{q}^{T}\textbf{B} + \textbf{r}^{T}\Gamma & \textbf{q}^{T}\Gamma^{T} + \textbf{r}^{T}\textbf{C}
\end{bmatrix}
\begin{bmatrix}
     \textbf{q} \\
     \textbf{r}
\end{bmatrix}
\]
\begin{equation}
    = (\textbf{q}^{T}\textbf{B}\textbf{q} + \textbf{r}^{T}\Gamma\textbf{q} + \textbf{q}^{T}\Gamma^{T}\textbf{r} + \textbf{r}^{T}\textbf{C}\textbf{r})
    = (\textbf{q}^{T}\textbf{B}\textbf{q} + 2\textbf{r}^{T}\Gamma\textbf{q} +  \textbf{r}^{T}\textbf{C}\textbf{r}). \nonumber
\end{equation}

\[ \text{Then, } \textbf{A}^{T}\textbf{w(t)} =
\begin{bmatrix}
     \textbf{R}^{T} & 0
\end{bmatrix} \textbf{P}^{T} \textbf{w(t)}
=
\begin{bmatrix}
     \textbf{R}^{T} & 0
\end{bmatrix} 
\begin{bmatrix}
     \textbf{q} \\
     \textbf{r}
\end{bmatrix}
= \textbf{b}
\]
\begin{equation}
   \text{or, } \textbf{R}^{T} \textbf{q} = \textbf{b} \nonumber
\end{equation}

\begin{equation}\label{eq1}
    \text{and finally, } \textbf{q} = \textbf{R}^{-T} \textbf{b}  
\end{equation}
The value of \textbf{q} helps to determine the term $ \textbf{q}^{T}\textbf{B}\textbf{q}$, so the objective function becomes $(2\textbf{r}^{T}\Gamma\textbf{q} +  \textbf{r}^{T}\textbf{C}\textbf{r})$ which now needs to be minimized. 
From the last constraint equation,
\[ \textbf{w(t)}^{T}\textbf{w(t)} = 
\begin{bmatrix}
   \textbf{q}^{T} & \textbf{r}^{T} 
\end{bmatrix}
\begin{bmatrix}
     \textbf{q} \\
     \textbf{r}
\end{bmatrix} =  \textbf{q}^{T} \textbf{q} +  \textbf{r}^{T} \textbf{r} = 1.
\]

We define
\begin{equation}
    s^{2} = 1 - \textbf{q}^{T} \textbf{q} = \textbf{r}^{T} \textbf{r} \nonumber
\end{equation}
\begin{equation}
    \text{and } \textbf{g} = - \Gamma \textbf{q} \nonumber
\end{equation}

So the optimization problem now becomes,

\begin{maxi}|l|
	  {\textbf{r}}{-2\textbf{r}^{T}\textbf{g} +  \textbf{r}^{T}\textbf{C}\textbf{r}}{}{}
	  \addConstraint{\textbf{r}^{T} \textbf{r}}{=s^{2}}.
	  \nonumber
\end{maxi}

But we can see that $2\textbf{r}^{T}\textbf{g}$ is a scalar quantity, therefore we can write $2 \textbf{r}^{T}\textbf{g} = 2 \textbf{g}^{T}\textbf{r}$. So the optimization function becomes,
\begin{maxi}|l|
	  {\textbf{r}}{- 2\textbf{g}^{T}\textbf{r} +  \textbf{r}^{T}\textbf{C}\textbf{r}}{}{}
	  \addConstraint{\textbf{r}^{T} \textbf{r}}{=s^{2}}.
	  \nonumber
\end{maxi}

Now using the Lagrangian multiplier we write the objective as
\begin{equation}\label{eq3}
    \phi(r, \lambda) = -2 \textbf{g}^{T}\textbf{r} +  \textbf{r}^{T}\textbf{C}\textbf{r} - \lambda ( \textbf{r}^{T} \textbf{r} - s^{2})
\end{equation}

Differentiating \eqref{eq3} with \textbf{r} and $\lambda $and equating to zero we get
\begin{equation}
- 2 \textbf{g} + 2 \textbf{C}\textbf{r} - 2 \lambda r = 0 \nonumber
\end{equation}

\begin{equation}
\text{such that } \textbf{r}^{T} \textbf{r} = s^{2}. \nonumber
\end{equation}

Normalizing the equations we get,
\begin{equation}\label{eq4}
\textbf{C}\textbf{r}  = \textbf{g} + \lambda r 
\end{equation}
\begin{equation}\label{eq5}
\text{and } \textbf{r}^{T} \textbf{r} = s^{2} 
\end{equation}

Doing an Eigenvalue decomposition of \textbf{C} we get $\textbf{C} = \textbf{Q}\textbf{D}\textbf{Q}^{T},$ 
where $\textbf{Q}^{T}\textbf{Q} = 1$ and $\textbf{D} = diag (\delta_1, \delta_2,....,\delta_{(n - 2)}).$

Now substituting it in equation \eqref{eq4} and \eqref{eq5} we obtain
\begin{equation}
\textbf{Q}\textbf{D}\textbf{Q}^{T}\textbf{r} = \textbf{g} + \lambda\textbf{Q}\textbf{Q}^{T}\textbf{r}. \nonumber
\end{equation}
Multiplying the entire equation by 
$\textbf{Q}^{T},$
\begin{equation}
\textbf{Q}^{T}\textbf{Q}\textbf{D}\textbf{Q}^{T}\textbf{r} = \textbf{Q}^{T}\textbf{g} + \textbf{Q}^{T}\lambda\textbf{Q}\textbf{Q}^{T}\textbf{r} \nonumber
\end{equation}
\begin{equation}
\text{and } \textbf{r}^{T}\textbf{Q}^{T}\textbf{Q}\textbf{r} = s^{2}. \nonumber
\end{equation}
As\quad $\textbf{Q}^{T}\textbf{Q} = 1$ therefore,
\begin{equation}\label{eq6}
\textbf{D}\textbf{Q}^{T}\textbf{r} = \textbf{Q}^{T}\textbf{g} + \lambda\textbf{Q}^{T}\textbf{r}. 
\end{equation}
Let us define
\begin{equation}
    \textbf{u} = \textbf{Q}^{T}\textbf{r}\nonumber
\end{equation}
\begin{equation}
\text{and } \textbf{d} = \textbf{Q}^{T}\textbf{g}\nonumber
\end{equation}

Thus equation \eqref{eq6} reduces to 
\begin{equation}\label{eq7}
\textbf{D}\textbf{u} = \textbf{d} + \lambda\textbf{u} 
\end{equation}
and 
\begin{equation} \label{eq8}
\textbf{u}^{T} \textbf{u} = s^{2}
\end{equation}

Solving equation \eqref{eq7} and \eqref{eq8},
we get 
$\textbf{u} = 
\begin{bmatrix}
    u_1 \\
    u_2 \\ u_3 \\ ....\\ u_{(n - 2)}
\end{bmatrix}.$
Once we get \textbf{u} we can get \textbf{r} as
$\textbf{r} = \textbf{Q}^{-T}\textbf{u}$
where $\textbf{r} = 
\begin{bmatrix}
    r_1 \\
    r_2 \\ r_3 \\ ....\\ r_{(n - 2)}
\end{bmatrix}.$
We already have 
$\textbf{q} = 
\begin{bmatrix}
    q_1 \\
    q_2 
\end{bmatrix},$ 
and  $\textbf{P}^{T}\textbf{w(t)} =
\begin{bmatrix}
      q_1 \\
    q_2 \\
    r_1 \\
    r_2 \\ r_3 \\ ....\\ r_{(n - 2)}
\end{bmatrix}$
and hence $\textbf{w(t)} = \textbf{P}
\begin{bmatrix}
      q_1 \\
    q_2 \\
    r_1 \\
    r_2 \\ r_3 \\ ....\\ r_{(n - 2)}
\end{bmatrix}.$

\textbf{Example: } Let us take an example for 3 stocks. For the simplification of algebraic calculation we take $\mu_1 = 0$.

\[ \text{Then } \Omega = 
\begin{bmatrix}
    \sigma_{1}^2(x_t) & \rho_{(12)}\sigma_{1}(x_t)\sigma_{2}(x_t) & \rho_{(13)}\sigma_{1}(x_t)\sigma_{3}(x_t)\\
    \rho_{(21)}\sigma_{2}(x_t)\sigma_{1}(x_t) & \sigma_{2}^2(x_t) & \rho_{(23)}\sigma_{2}(x_t)\sigma_{3}(x_t)\\
    \rho_{(31)}\sigma_{3}(x_t)\sigma_{1}(x_t) & \rho_{(32)}\sigma_{3}(x_t)\sigma_{2}(x_t) & \sigma_{3}^2(x_t)
\end{bmatrix}
\]

\[ \text{and }
\textbf{A}=
\begin{bmatrix}
    0 & 1 \\
    \mu_2 & 1\\
    \mu_3 & 1 \\
\end{bmatrix}.
\]

Doing a QR decomposition of A, we get
\[
\textbf{P} = \begin{bmatrix}
\textbf{P}_1 & \textbf{P}_2
\end{bmatrix},
\]
\[ \text{where }
\textbf{P}_1=
\begin{bmatrix}
    0 & \frac{\mu^2_2 + \mu^2_3}{\sqrt{2(\mu^2_2 + \mu^2_3 - \mu_2\mu_3)(\mu^2_2 + \mu^2_3)}} \\
   \frac{\mu_2}{\sqrt{\mu^2_2 + \mu^2_3}} & \frac{ \mu^2_3 - \mu_2\mu_3}{\sqrt{2(\mu^2_2 + \mu^2_3 - \mu_2\mu_3)(\mu^2_2 + \mu^2_3)}}\\
   \frac{\mu_3}{\sqrt{\mu^2_2 + \mu^2_3}} & \frac{\mu^2_2 - \mu_3\mu_2}{\sqrt{2(\mu^2_2 + \mu^2_3 - \mu_2\mu_3)(\mu^2_2 + \mu^2_3)}}
   
\end{bmatrix}
\]

\[ \text{and }
\textbf{P}_2=
\begin{bmatrix}
     \frac{\mu_2 - \mu_3}{\sqrt{2(\mu^2_2 + \mu^2_3 - \mu_2\mu_3)}} \\
   \frac{\mu_3}{\sqrt{2(\mu^2_2 + \mu^2_3 - \mu_2\mu_3)}}\\
   -\frac{\mu_2}{\sqrt{2(\mu^2_2 + \mu^2_3 - \mu_2\mu_3)}}
   
\end{bmatrix}.
\]

\[\textbf{A} = 
\begin{bmatrix}
\textbf{P}_1 & \textbf{P}_2
\end{bmatrix}
\begin{bmatrix}
\textbf{R} \\ 0
\end{bmatrix}
\]

\[ \text{where }
\textbf{R}=
\begin{bmatrix}
    \sqrt{\mu^2_2 + \mu^2_3} & \frac{\mu_2 + \mu_3}{\sqrt{\mu^2_2 + \mu^2_3}} \\
   0 & \sqrt{\frac{2(\mu^2_2 + \mu^2_3 - \mu_2\mu_3 )}{\mu^2_2 + \mu^2_3}}\\
\end{bmatrix}.
\]

We are going to calculate \textbf{B}, $\Gamma$ and \textbf{C} from the below equations as previously defined
\[
\textbf{R}^{-1} = \frac{1}{\det R}
\begin{bmatrix}
     \sqrt{\frac{2(\mu^2_2 + \mu^2_3 - \mu_2\mu_3)}{\mu^2_2 + \mu^2_3}} & -\frac{\mu_2 + \mu_3}{\sqrt{\mu^2_2 + \mu^2_3}} \\
   0 & \sqrt{\mu^2_2 + \mu^2_3}\\
\end{bmatrix}
\]
\[
\text{or } \textbf{R}^{-T} = \frac{1}{\det R}
\begin{bmatrix}
    \sqrt{\frac{2(\mu^2_2 + \mu^2_3 - \mu_2\mu_3)}{\mu^2_2 + \mu^2_3}} & 0 \\
   -\frac{\mu_2 + \mu_3}{\sqrt{\mu^2_2 + \mu^2_3}} & \sqrt{\mu^2_2 + \mu^2_3}\\
\end{bmatrix}
\]

\[
\text{and } \textbf{q} = \frac{1}{\det R}
\begin{bmatrix}
    \sqrt{\frac{2(\mu^2_2 + \mu^2_3 - \mu_2\mu_3)}{\mu^2_2 + \mu^2_3}} & 0 \\
   -\frac{\mu_2 + \mu_3}{\sqrt{\mu^2_2 + \mu^2_3}} & \sqrt{\mu^2_2 + \mu^2_3}\\
\end{bmatrix}
\begin{bmatrix}
    k \\
   1
\end{bmatrix}
\]
\[ \text{or }
\textbf{q} = \frac{1}{\det R}
\begin{bmatrix}
     k \sqrt{\frac{2(\mu^2_2 + \mu^2_3 - \mu_2\mu_3)}{\mu^2_2 + \mu^2_3}} \\
   -k\big(\frac{\mu_2 + \mu_3}{\sqrt{\mu^2_2 + \mu^2_3}}\big) + \sqrt{\mu^2_2 + \mu^2_3}
\end{bmatrix}.
\]
We finally want to compute
\[
\begin{bmatrix}
\textbf{P}_1 & \textbf{P}_2
\end{bmatrix}^{T} \begin{bmatrix}
    \sigma_{1}^2(x_t) & \rho_{(12)}\sigma_{1}(x_t)\sigma_{2}(x_t) & \rho_{(13)}\sigma_{1}(x_t)\sigma_{3}(x_t)\\
    \rho_{(21)}\sigma_{2}(x_t)\sigma_{1}(x_t) & \sigma_{2}^2(x_t) & \rho_{(23)}\sigma_{2}(x_t)\sigma_{3}(x_t)\\
    \rho_{(31)}\sigma_{3}(x_t)\sigma_{1}(x_t) & \rho_{(32)}\sigma_{3}(x_t)\sigma_{2}(x_t) & \sigma_{3}^2(x_t)
\end{bmatrix}
\begin{bmatrix}
\textbf{P}_1 & \textbf{P}_2
\end{bmatrix}.
\]

Let us consider the following definitions,

$\sqrt{\mu^2_2 + \mu^2_3} = Y, $ \newline
$\sqrt{2(\mu^2_2 + \mu^2_3 - \mu_2\mu_3)(\mu^2_2 + \mu^2_3)} = X,$ \newline
$\sqrt{2(\mu^2_2 + \mu^2_3 - \mu_2\mu_3)} = Z,$\newline
and $\mu_2 - \mu_3 = V.$

As \textbf{r} contains only 1 element, we can use the definition 
\begin{equation}
    s^{2} = 1 - \textbf{q}^{T} \textbf{q} = \textbf{r}^{T} \textbf{r} \nonumber
\end{equation}
to calculate \textbf{r}
\[
\textbf{w(t)} = \textbf{P}
\begin{bmatrix}
     \frac{1}{\det R}\big(k \sqrt{\frac{2(\mu^2_2 + \mu^2_3 - \mu_2\mu_3)}{\mu^2_2 + \mu^2_3}}\big) \\
   \frac{1}{\det R}\Big(-k\big(\frac{\mu_2 + \mu_3}{\sqrt{\mu^2_2 + \mu^2_3}}\big) + \sqrt{\mu^2_2 + \mu^2_3} \Big) \\ \textbf{r}
\end{bmatrix}
\]
Using the notations
\[
\textbf{w(t)} = \textbf{P}
\begin{bmatrix}
     \frac{1}{\det R}\big(\frac{kZ}{Y}\big) \\
   \frac{1}{\det R}\big(-k\big(\frac{\mu_2 + \mu_3}{Y}\big) + Y\big) \\ \textbf{r}
\end{bmatrix}
\]

\[
\textbf{r}^{T}\textbf{r} = 1 - \Bigg(\frac{1}{\det R}
\begin{bmatrix}
     k \sqrt{\frac{2(\mu^2_2 + \mu^2_3 - \mu_2\mu_3)}{\mu^2_2 + \mu^2_3}} \\
   -k\big(\frac{\mu_2 + \mu_3}{\sqrt{\mu^2_2 + \mu^2_3}}\big) + \sqrt{\mu^2_2 + \mu^2_3}
\end{bmatrix}^{T}
\frac{1}{\det R}
\begin{bmatrix}
     k \sqrt{\frac{2(\mu^2_2 + \mu^2_3 - \mu_2\mu_3)}{\mu^2_2 + \mu^2_3}} \\
   -k\big(\frac{\mu_2 + \mu_3}{\sqrt{\mu^2_2 + \mu^2_3}}\big) + \sqrt{\mu^2_2 + \mu^2_3}
\end{bmatrix}\Bigg)
\]

\begin{eqnarray}
   \textbf{r} = \sqrt{1 - \frac{1}{(\det R)^2} \Big(k^2 \frac{2(\mu^2_2 + \mu^2_3 - \mu_2\mu_3)}{\mu^2_2 + \mu^2_3} + 
  \big(\big(-k\frac{\mu_2 + \mu_3}{\sqrt{\mu^2_2 + \mu^2_3}}\big) + \sqrt{\mu^2_2 + \mu^2_3}\big)^2\Big)} \nonumber
\end{eqnarray}
\begin{eqnarray}
   \textbf{r} = \sqrt{1 - \frac{1}{(\det R)^2} \Big(k^2 \frac{Z^2}{Y^2} + 
  \big(\big(-k\frac{\mu_2 + \mu_3}{Y}\big) + Y\big)^2\Big)} \nonumber
\end{eqnarray}
Therefore we can calculate \textbf{w(t)}
\[
\textbf{w(t)} = \textbf{P}
\begin{bmatrix}
      \frac{1}{\det R}\big(\frac{k Z}{Y}\big) \\
   \frac{1}{\det R}\big(-k\big(\frac{\mu_2 + \mu_3}{Y}\big) + Y\big) \\ \sqrt{1 - \frac{1}{(\det R)^2} \big(k^2 \frac{Z^2}{Y^2} + 
  \big(\big(-k\frac{\mu_2 + \mu_3}{Y}\big) + Y\big)^2\big)}
\end{bmatrix}
\]

We can denote 
\[\textbf{F} = 
\begin{bmatrix}
      \frac{1}{\det R}\big(\frac{k Z}{Y}\big) \\
   \frac{1}{\det R}\big(-k\big(\frac{\mu_2 + \mu_3}{Y}\big) + Y\big) \\ \sqrt{1 - \frac{1}{(\det R)^2} \big(k^2 \frac{Z^2}{Y^2} + 
  \big(\big(-k\frac{\mu_2 + \mu_3}{Y}\big) + Y\big)^2\big)}
\end{bmatrix}
\]
So, \textbf{w(t)} becomes $\textbf{w(t)} = \textbf{P}\textbf{F}$. We can write the maximum eigenvalue as,

\begin{equation}
    \lambda(x_t) = \textbf{w(t)}^{T}\Omega \textbf{w(t)}\nonumber
\end{equation}
\begin{equation}
    \lambda(x_t) = \textbf{F}^{T}\textbf{P}^{T}\Omega\textbf{P}\textbf{F}\nonumber
\end{equation}

We can now use the following definitions
\[\textbf{P}^{T}\Omega\textbf{P} =
\begin{bmatrix}
     \textbf{B} & \Gamma^{T} \\
     \Gamma & \textbf{C}
\end{bmatrix} = 
\begin{bmatrix}
     \textbf{k}_1 & \textbf{k}_2 &
     \textbf{k}_3
\end{bmatrix}
\]

With the previous definition we can simplify \textbf{P} as
\[
\textbf{P}=
\begin{bmatrix}
    0 & \frac{\mu^2_2 + \mu^2_3}{X} & \frac{\mu_2 - \mu_3}{Z} \\
   \frac{\mu_2}{Y} & \frac{\mu^2_3 - \mu_2\mu_3}{X} &  \frac{\mu_3}{Z}\\
   \frac{\mu_3}{Y} & \frac{\mu^2_2 - \mu_3\mu_2}{X} & -\frac{\mu_2}{Z}
\end{bmatrix} = 
\begin{bmatrix}
    0 & \frac{Y^2}{X} & \frac{V}{Z} \\
   \frac{\mu_2}{Y} & -\frac{V\mu_3}{X} &  \frac{\mu_3}{Z}\\
   \frac{\mu_3}{Y} & \frac{V\mu_2}{X} & -\frac{\mu_2}{Z}
\end{bmatrix}
\]
Then, $\textbf{P}^{T}\Omega\textbf{P} = $
\[
\begin{bmatrix}
    0 & \frac{Y^2}{X} & \frac{V}{Z} \\
   \frac{\mu_2}{Y} & -\frac{V\mu_3}{X} &  \frac{\mu_3}{Z}\\
   \frac{\mu_3}{Y} & \frac{V\mu_2}{X} & -\frac{\mu_2}{Z}
\end{bmatrix}^{T}
\begin{bmatrix}
    \sigma_{1}^2(x_t) & \rho_{(12)}\sigma_{1}(x_t)\sigma_{2}(x_t) & \rho_{(13)}\sigma_{1}(x_t)\sigma_{3}(x_t)\\
    \rho_{(21)}\sigma_{2}(x_t)\sigma_{1}(x_t) & \sigma_{2}^2(x_t) & \rho_{(23)}\sigma_{2}(x_t)\sigma_{3}(x_t)\\
    \rho_{(31)}\sigma_{3}(x_t)\sigma_{1}(x_t) & \rho_{(32)}\sigma_{3}(x_t)\sigma_{2}(x_t) & \sigma_{3}^2(x_t)
\end{bmatrix}
\begin{bmatrix}
    0 & \frac{Y^2}{X} & \frac{V}{Z} \\
   \frac{\mu_2}{Y} & -\frac{V\mu_3}{X} &  \frac{\mu_3}{Z}\\
   \frac{\mu_3}{Y} & \frac{V\mu_2}{X} & -\frac{\mu_2}{Z}
\end{bmatrix}
\]
Multiplying out, we get the individual vectors $\textbf{k}_1$ etc. as
\[\textbf{k}_1 =
\begin{bmatrix}
    \frac{\mu_2}{Y}\big(\frac{\mu_2}{Y}\sigma^2_2(x_t) + \frac{\mu_3}{Y}\rho_{32}\sigma_3(x_t)\sigma_2(x_t)\big) + \frac{\mu_3}{Y}\big(\frac{\mu_2}{Y}\rho_{23}\sigma_2(x_t)\sigma_3(x_t) + \frac{\mu_3}{Y}\sigma^2_3(x_t)\big) \\[2.5ex] \frac{\mu_2}{Y}\big(\frac{Y^2}{X}\rho_{12}\sigma_1(x_t)\sigma_2(x_t) - \frac{V\mu_3}{X}\sigma^2_2(x_t) + \frac{V\mu_2}{X}\rho_{32}\sigma_3(x_t)\sigma_2(x_t)\big) + \frac{\mu_3}{Y}\big(\frac{Y^2}{X}\rho_{13}\sigma_1(x_t)\sigma_3(x_t) - \frac{V\mu_3}{X}\rho_{23}\sigma_2(x_t)\sigma_3(x_t) + \frac{V\mu_2}{X}\sigma^2_3(x_t)\big) \\[2.5ex]  \frac{\mu_2}{Y}\big(\frac{V}{Z}\rho_{12}\sigma_1(x_t)\sigma_2(x_t) + \frac{\mu_3}{Z}\sigma^2_2(x_t) - \frac{\mu_2}{Z}\rho_{32}\sigma_3(x_t)\sigma_2(x_t)\big) + \frac{\mu_3}{Y}\big(\frac{V}{Z}\rho_{13}\sigma_1(x_t)\sigma_3(x_t) + \frac{\mu_3}{Z}\rho_{23}\sigma_2(x_t)\sigma_3(x_t) - \frac{\mu_2}{Z}\sigma^2_3(x_t)\big) 
\end{bmatrix}
\]
\[ \textbf{k}_2 = 
 \begin{bmatrix}
 \begin{matrix}
  \frac{Y^2}{X}\big(\frac{\mu_2}{Y}\rho_{21}\sigma_2(x_t)\sigma_1(x_t) + \frac{\mu_3}{Y}\rho_{31}\sigma_3(x_t)\sigma_1(x_t)\big) - \\ \hfill{} \frac{V\mu_3}{X}\big(\frac{\mu_2}{Y}\sigma^2_2(x_t) + \frac{\mu_3}{Y}\rho_{32}\sigma_3(x_t)\sigma_2(x_t)\big) + \frac{V\mu_2}{X}\big(\frac{\mu_2}{Y}\rho_{23}\sigma_2(x_t)\sigma_3(x_t) + \frac{\mu_3}{Y}\sigma^2_3(x_t)\big) 
  \end{matrix}
  \\[5ex] 
  \begin{matrix}
  \frac{Y^2}{x}\big(\frac{Y^2}{X}\sigma^2_1(x_t) - \frac{V\mu_3}{X}\rho_{21}\sigma_2(x_t)\sigma_1(x_t) + \frac{V\mu_2}{X}\rho_{31}\sigma_3(x_t)\sigma_1(x_t)\big) - \\ \hfill{} \frac{V\mu_3}{X}\big(\frac{Y^2}{X}\rho_{12}\sigma_1(x_t)\sigma_2(x_t) - \frac{V\mu_3}{X}\sigma^2_2(x_t) + \frac{V\mu_2}{X}\rho_{32}\sigma_3(x_t)\sigma_2(x_t)\big) + \frac{V\mu_2}{X}\big(\frac{Y^2}{X}\rho_{13}\sigma_1(x_t)\sigma_3(x_t) - \frac{V\mu_3}{X}\rho_{23}\sigma_2(x_t)\sigma_3(x_t) + \frac{V\mu_2}{X}\sigma^2_3(x_t)\big)
  \end{matrix}
  \\[5ex]
  \begin{matrix}
    \frac{V}{Z}\big(\frac{V}{Z}\sigma^2_1(x_t) - \frac{V\mu_3}{X}\rho_{21}\sigma_2(x_t)\sigma_1(x_t) + \frac{V\mu_2}{X}\rho_{31}\sigma_3(x_t)\sigma_1(x_t)\big) + \\ \hfill{} \frac{\mu_3}{Z}\big(\frac{V}{Z}\rho_{12}\sigma_1(x_t)\sigma_2(x_t) + \frac{\mu_3}{Z}\sigma^2_2(x_t) - \frac{\mu_2}{Z}\rho_{32}\sigma_3(x_t)\sigma_2(x_t)\big) - \frac{\mu_2}{Z}\big(\frac{V}{Z}\rho_{13}\sigma_1(x_t)\sigma_3(x_t) + \frac{\mu_3}{Z}\rho_{23}\sigma_2(x_t)\sigma_3(x_t) - \frac{\mu_2}{Z}\sigma^3_3(x_t)\big) 
    \end{matrix}
    \end{bmatrix}
    \]
 \[ \textbf{k}_3 = 
    \begin{bmatrix}
\begin{matrix}
\frac{V}{Z}\big(\frac{\mu_2}{Y}\rho_{21}\sigma_2(x_t)\sigma_1(x_t) + \frac{\mu_3}{Y}\rho_{31}\sigma_3(x_t)\sigma_1(x_t)\big)+\\ \hfill{}  \frac{\mu_3}{Z}\big(\frac{\mu_2}{Y}\sigma^2_2(x_t) + \frac{\mu_3}{Y}\rho_{32}\sigma_3(x_t)\sigma_2(x_t)\big)  - \frac{\mu_2}{X}\big(\frac{\mu_2}{Y}\rho_{23}\sigma_2(x_t)\sigma_3(x_t) + \frac{\mu_3}{Y}\sigma^2_3(x_t)\big)   \end{matrix}        \\[5ex]
\begin{matrix}
\frac{V}{Z}\big(\frac{Y^2}{X}\sigma^2_1(x_t) - \frac{V\mu_3}{X}\rho_{21}\sigma_2(x_t)\sigma_1(x_t) + \frac{V\mu_2}{X}\rho_{31}\sigma_3(x_t)\sigma_1(x_t)\big) +\\ \hfill{}  \frac{\mu_3}{Z}\big(\frac{Y^2}{X}\rho_{12}\sigma_1(x_t)\sigma_2(x_t) - \frac{V\mu_3}{X}\sigma^2_2(x_t) + \frac{V\mu_2}{X}\rho_{32}\sigma_3(x_t)\sigma_2(x_t)\big) - \frac{\mu_2}{Z}\big(\frac{Y^2}{X}\rho_{13}\sigma_1(x_t)\sigma_3(x_t) - \frac{V\mu_3}{X}\rho_{23}\sigma_2(x_t)\sigma_3(x_t) + \frac{V\mu_2}{X}\sigma^2_3(x_t)\big) 
\end{matrix}\\[5ex]
\begin{matrix}
        \frac{V}{Z}\big(\frac{V}{Z}\sigma^2_1(x_t) + \frac{\mu_3}{Z}\rho_{21}\sigma_2(x_t)\sigma_1(x_t) - \frac{\mu_2}{Z}\rho_{31}\sigma_3(x_t)\sigma_1(x_t)\big) + \\ \hfill{} \frac{\mu_3}{Z}\big(\frac{V}{Z}\rho_{12}\sigma_1(x_t)\sigma_2(x_t) + \frac{\mu_3}{Z}\sigma^2_2(x_t) - \frac{\mu_2}{Z}\rho_{32}\sigma_3(x_t)\sigma_2(x_t)\big) - \frac{\mu_2}{Z}\big(\frac{Y^2}{X}\rho_{13}\sigma_1(x_t)\sigma_3(x_t) + \frac{\mu_3}{Z}\rho_{23}\sigma_2(x_t)\sigma_3(x_t) - \frac{\mu_2}{z}\sigma^2_3(x_t)\big)
 \end{matrix}  
\end{bmatrix}
\]
\\
So, we can calculate the eigenvalue as, 
\[ \lambda(x_t) = 
\textbf{F}^{T}  
\begin{bmatrix}
 \textbf{k}_1 & \textbf{k}_2 & \textbf{k}_3   
\end{bmatrix}
\textbf{F}
\]

where 
\[
\textbf{F}^{T}  
\begin{bmatrix}
 \textbf{k}_1 & \textbf{k}_2 & \textbf{k}_3   
\end{bmatrix} = 
\begin{bmatrix}
 \textbf{d}_1 & \textbf{d}_2 & \textbf{d}_3   
\end{bmatrix}
\]
\begin{eqnarray}
  \textbf{d}_1 &=& \Big(\frac{\mu_2}{Y}\big(\frac{\mu_2}{Y}\sigma^2_2(x_t) + \frac{\mu_3}{Y}\rho_{32}\sigma_3(x_t)\sigma_2(x_t)\big) + \frac{\mu_3}{Y}\big(\frac{\mu_2}{Y}\rho_{23}\sigma_2(x_t)\sigma_3(x_t) + \frac{\mu_3}{Y}\sigma^2_3(x_t)\big)\Big)\Big(\frac{1}{\det R}\big(\frac{k Z}{Y}\big)\Big) + \nonumber \\&& 
  \Big(\frac{\mu_2}{Y}\big(\frac{Y^2}{X}\rho_{12}\sigma_1(x_t)\sigma_2(x_t) - \frac{V\mu_3}{X}\sigma^2_2(x_t) + \frac{V\mu_2}{X}\rho_{32}\sigma_3(x_t)\sigma_2(x_t)\big) + \frac{\mu_3}{Y}\big(\frac{Y^2}{X}\rho_{13}\sigma_1(x_t)\sigma_3(x_t) - \frac{V\mu_3}{X}\rho_{23}\sigma_2(x_t)\sigma_3(x_t) + \nonumber \\&&  \frac{V\mu_2}{X}\sigma^2_3(x_t)\big)\Big)\Big(\frac{1}{\det R}\big(-k\big(\frac{\mu_2 + \mu_3}{Y}\big) + Y\big)\Big) + \nonumber \\&&
  \Big(\frac{\mu_2}{Y}\big(\frac{V}{Z}\rho_{12}\sigma_1(x_t)\sigma_2(x_t) + \frac{\mu_3}{Z}\sigma^2_2(x_t) - \frac{\mu_2}{Z}\rho_{32}\sigma_3(x_t)\sigma_2(x_t)\big) + \frac{\mu_3}{Y}\big(\frac{V}{Z}\rho_{13}\sigma_1(x_t)\sigma_3(x_t) + \frac{\mu_3}{Z}\rho_{23}\sigma_2(x_t)\sigma_3(x_t) - \nonumber \\&& \frac{\mu_2}{Z}\sigma^2_3(x_t)\big)\Big) \sqrt{1 - \frac{1}{(\det R)^2} \Big(k^2 \frac{Z^2}{Y^2} + 
  \big(\big(-k\frac{\mu_2 + \mu_3}{Y}\big) + Y\big)^2\Big)} \nonumber
  \end{eqnarray}

\begin{eqnarray}
  \textbf{d}_2 &=& 
  \Big(\frac{Y^2}{X}\big(\frac{\mu_2}{Y}\rho_{21}\sigma_2(x_t)\sigma_1(x_t) + \frac{\mu_3}{Y}\rho_{31}\sigma_3(x_t)\sigma_1(x_t)\big) - \nonumber \\&& \frac{V\mu_3}{X}\big(\frac{\mu_2}{Y}\sigma^2_2(x_t) + \frac{\mu_3}{Y}\rho_{32}\sigma_3(x_t)\sigma_2(x_t)\big) + \frac{V\mu_2}{X}\big(\frac{\mu_2}{Y}\rho_{23}\sigma_2(x_t)\sigma_3(x_t) +  \frac{\mu_3}{Y}\sigma^2_3(x_t)\big)\Big)\Big(\frac{1}{\det R}\big(\frac{k Z}{Y}\big)\Big) + \nonumber \\&& \Big(\frac{Y^2}{x}\big(\frac{Y^2}{X}\sigma^2_1(x_t) - \frac{V\mu_3}{X}\rho_{21}\sigma_2(x_t)\sigma_1(x_t) + \frac{V\mu_2}{X}\rho_{31}\sigma_3(x_t)\sigma_1(x_t)\big) - \frac{V\mu_3}{X}\big(\frac{Y^2}{X}\rho_{12}\sigma_1(x_t)\sigma_2(x_t) - \frac{V\mu_3}{X}\sigma^2_2(x_t) + \nonumber \\&& \frac{V\mu_2}{X}\rho_{32}\sigma_3(x_t)\sigma_2(x_t)\big) + \frac{V\mu_2}{X}\big(\frac{Y^2}{X}\rho_{13}\sigma_1(x_t)\sigma_3(x_t) - \frac{V\mu_3}{X}\rho_{23}\sigma_2(x_t)\sigma_3(x_t) + \frac{V\mu_2}{X}\sigma^2_3(x_t)\big)\Big)\Big(\frac{1}{\det R}\big(-k\big(\frac{\mu_2 + \mu_3}{Y}\big) + Y\big)\Big) + \nonumber \\&&
   \Big(\frac{V}{Z}\big(\frac{V}{Z}\sigma^2_1(x_t) - \frac{V\mu_3}{X}\rho_{21}\sigma_2(x_t)\sigma_1(x_t) + \frac{V\mu_2}{X}\rho_{31}\sigma_3(x_t)\sigma_1(x_t)\big) +  \frac{\mu_3}{Z}\big(\frac{V}{Z}\rho_{12}\sigma_1(x_t)\sigma_2(x_t) + \frac{\mu_3}{Z}\sigma^2_2(x_t) -  \frac{\mu_2}{Z}\rho_{32}\sigma_3(x_t)\sigma_2(x_t)\big) \nonumber \\&& - \frac{\mu_2}{Z}\big(\frac{V}{Z}\rho_{13}\sigma_1(x_t)\sigma_3(x_t) + \frac{\mu_3}{Z}\rho_{23}\sigma_2(x_t)\sigma_3(x_t) - \frac{\mu_2}{Z}\sigma^2_3(x_t)\big)\Big)\sqrt{1 - \frac{1}{(\det R)^2} \Big(k^2 \frac{Z^2}{Y^2} + 
  \big(\big(-k\frac{\mu_2 + \mu_3}{Y}\big) + Y\big)^2\Big)} \nonumber
\end{eqnarray}

\begin{eqnarray}
   \textbf{d}_3 &=&
   \Big(\frac{V}{Z}\big(\frac{\mu_2}{Y}\rho_{21}\sigma_2(x_t)\sigma_1(x_t) + \frac{\mu_3}{Y}\rho_{31}\sigma_3(x_t)\sigma_1(x_t)\big)+ \nonumber \\&&  \frac{\mu_3}{Z}\big(\frac{\mu_2}{Y}\sigma^2_2(x_t) + \frac{\mu_3}{Y}\rho_{32}\sigma_3(x_t)\sigma_2(x_t)\big)  - \frac{\mu_2}{X}\big(\frac{\mu_2}{Y}\rho_{23}\sigma_2(x_t)\sigma_3(x_t) + \frac{\mu_3}{Y}\sigma^2_3(x_t)\big)\Big)\Big(\frac{1}{\det R}\big(\frac{k Z}{Y}\big)\Big) + \nonumber \\&&
   \Big(\frac{V}{Z}\big(\frac{Y^2}{X}\sigma^2_1(x_t) - \frac{V\mu_3}{X}\rho_{21}\sigma_2(x_t)\sigma_1(x_t) + \frac{V\mu_2}{X}\rho_{31}\sigma_3(x_t)\sigma_1(x_t)\big) +  \frac{\mu_3}{Z}\big(\frac{Y^2}{X}\rho_{12}\sigma_1(x_t)\sigma_2(x_t) - \frac{V\mu_3}{X}\sigma^2_2(x_t) + \nonumber \\&& \frac{V\mu_2}{X}\rho_{32}\sigma_3(x_t)\sigma_2(x_t)\big) - \frac{\mu_2}{Z}\big(\frac{Y^2}{X}\rho_{13}\sigma_1(x_t)\sigma_3(x_t) - \frac{V\mu_3}{X}\rho_{23}\sigma_2(x_t)\sigma_3(x_t) + \frac{V\mu_2}{X}\sigma^2_3(x_t)\big)\Big)\Big(\frac{1}{\det R}\big(-k\big(\frac{\mu_2 + \mu_3}{Y}\big) + Y\big)\Big) + \nonumber \\&&
    \Big(\frac{V}{Z}\big(\frac{V}{Z}\sigma^2_1(x_t) + \frac{\mu_3}{Z}\rho_{21}\sigma_2(x_t)\sigma_1(x_t) - \frac{\mu_2}{Z}\rho_{31}\sigma_3(x_t)\sigma_1(x_t)\big) +  \frac{\mu_3}{Z}\big(\frac{V}{Z}\rho_{12}\sigma_1(x_t)\sigma_2(x_t) + \frac{\mu_3}{Z}\sigma^2_2(x_t) - \frac{\mu_2}{Z}\rho_{32}\sigma_3(x_t)\sigma_2(x_t)\big) - \nonumber \\&& \frac{\mu_2}{Z}\big(\frac{Y^2}{X}\rho_{13}\sigma_1(x_t)\sigma_3(x_t) +  \frac{\mu_3}{Z}\rho_{23}\sigma_2(x_t)\sigma_3(x_t) - \frac{\mu_2}{z}\sigma^2_3(x_t)\big)\Big)\sqrt{1 - \frac{1}{(\det R)^2} \Big(k^2 \frac{Z^2}{Y^2} +
  \big(\big(-k\frac{\mu_2 + \mu_3}{Y}\big) +  Y\big)^2\Big)} \nonumber
\end{eqnarray}

\begin{eqnarray}
   \lambda(x_t) = \textbf{d}_1 \Big(\frac{1}{\det R}\big(\frac{k Z}{Y}\big)\Big) + \textbf{d}_2 \Big(\frac{1}{\det R}\big(-k\big(\frac{\mu_2 + \mu_3}{Y}\big) + Y\big)\Big) + \textbf{d}_3 \sqrt{1 - \frac{1}{(\det R)^2} \Big(k^2 \frac{Z^2}{Y^2} + \big(\big(-k\frac{\mu_2 + \mu_3}{Y}\big) +  Y\big)^2\Big)} \nonumber
\end{eqnarray}
Now the price of the largest eigenvalue swap is the expected present value of the payoff in the risk neutral world for this three asset we have considered

\begin{equation} \label{eq9}
    P (x) = E \{e^{-rT}(\lambda (x_t) - K)\} 
\end{equation}
\begin{equation} 
    P (x) = e^{-rT} E \{(\lambda (x_t) - K)\} \nonumber
\end{equation}

\begin{eqnarray}
  P (x_1) &=& \Bigg( \frac{\mu_2}{Y}(\frac{\mu_2}{Y}\big(e^{-rT}\{\frac{1}{T}\int_{0}^{T}(e^{tQ} \sigma_{2}^2(x)) dt\}\big) + \frac{\mu_3}{Y}\big(e^{-rT}\{\frac{1}{T}\int_{0}^{T}(e^{tQ}(\rho_{32}\sigma_{3}(x)\sigma_{2}(x))dt\}\big)) + \nonumber \\&& \frac{\mu_3}{Y}(\frac{\mu_2}{Y}\big(e^{-rT}\{\frac{1}{T}\int_{0}^{T}(e^{tQ}(\rho_{23}\sigma_{2}(x)\sigma_{3}(x))dt\}\big) + \frac{\mu_3}{Y}\big(e^{-rT}\{\frac{1}{T}\int_{0}^{T}(e^{tQ} \sigma_{3}^2(x)) dt\}\big))\big(\frac{1}{\det R}\big(\frac{k Z}{Y}\big)\big) + \nonumber \\&& 
  \frac{\mu_2}{Y}(\frac{Y^2}{X}\big(e^{-rT}\{\frac{1}{T}\int_{0}^{T}(e^{tQ}(\rho_{12}\sigma_{1}(x)\sigma_{2}(x))dt\}\big)) - \frac{V\mu_3}{X}\big(e^{-rT}\{\frac{1}{T}\int_{0}^{T}(e^{tQ} \sigma_{2}^2(x)) dt\}\big) + \nonumber \\&& \frac{V\mu_2}{X}\big(e^{-rT}\{\frac{1}{T}\int_{0}^{T}(e^{tQ}(\rho_{32}\sigma_{3}(x)\sigma_{2}(x))dt\}\big))) +  \frac{\mu_3}{Y}(\frac{Y^2}{X}\big(e^{-rT}\{\frac{1}{T}\int_{0}^{T}(e^{tQ}(\rho_{13}\sigma_{1}(x)\sigma_{3}(x))dt\}\big)) - \nonumber \\&& \frac{V\mu_3}{X}\big(e^{-rT}\{\frac{1}{T}\int_{0}^{T}(e^{tQ}(\rho_{23}\sigma_{2}(x)\sigma_{3}(x))dt\}\big)) + \frac{V\mu_2}{X}\big(e^{-rT}\{\frac{1}{T}\int_{0}^{T}(e^{tQ} \sigma_{3}^2(x)) dt\}\big))\big(\frac{1}{\det R}\big(-k\big(\frac{\mu_2 + \mu_3}{Y}\big) + Y\big)\big) \nonumber \\&& + 
  \frac{\mu_2}{Y}(\frac{V}{Z}\big(e^{-rT}\{\frac{1}{T}\int_{0}^{T}(e^{tQ}(\rho_{12}\sigma_{1}(x)\sigma_{2}(x))dt\}\big)) + \frac{\mu_3}{Z}\big(e^{-rT}\{\frac{1}{T}\int_{0}^{T}(e^{tQ} \sigma_{2}^2(x)) dt\}\big) - \nonumber \\&& \frac{\mu_2}{Z}\big(e^{-rT}\{\frac{1}{T}\int_{0}^{T}(e^{tQ}(\rho_{32}\sigma_{3}(x)\sigma_{2}(x))dt\}\big))) + \frac{\mu_3}{Y}(\frac{V}{Z}\big(e^{-rT}\{\frac{1}{T}\int_{0}^{T}(e^{tQ}(\rho_{13}\sigma_{1}(x)\sigma_{3}(x))dt\}\big) + \nonumber \\&& \frac{\mu_3}{Z}\big(e^{-rT}\{\frac{1}{T}\int_{0}^{T}(e^{tQ}(\rho_{23}\sigma_{2}(x)\sigma_{3}(x))dt\}\big)) - \frac{\mu_2}{Z}\big(e^{-rT}\{\frac{1}{T}\int_{0}^{T}(e^{tQ} \sigma_{3}^2(x)) dt\}\big))  \nonumber \\&& \sqrt{1 - \frac{1}{(\det R)^2} \big(k^2 \frac{Z^2}{Y^2} + 
  \big(\big(-k\frac{\mu_2 + \mu_3}{Y}\big) + Y\big)^2\big)} \Bigg) \times \big(\frac{1}{\det R}\big(\frac{k Z}{Y}\big)\big) \nonumber    
\end{eqnarray}

\begin{eqnarray}
  P (x_2) &=& 
  \Bigg(\frac{Y^2}{X}((\frac{\mu_2}{Y}\big(e^{-rT}\{\frac{1}{T}\int_{0}^{T}(e^{tQ}(\rho_{21}\sigma_{2}(x)\sigma_{1}(x))dt\}\big) + \frac{\mu_3}{Y}\big(e^{-rT}\{\frac{1}{T}\int_{0}^{T}(e^{tQ}(\rho_{31}\sigma_{3}(x)\sigma_{1}(x))dt\}\big)) - \nonumber \\&& \frac{V\mu_3}{X}(\frac{\mu_2}{Y}\big(e^{-rT}\{\frac{1}{T}\int_{0}^{T}(e^{tQ} \sigma_{2}^2(x)) dt\}\big) + \frac{\mu_3}{Y}\big(e^{-rT}\{\frac{1}{T}\int_{0}^{T}(e^{tQ}(\rho_{32}\sigma_{3}(x)\sigma_{2}(x))dt\}\big)) + \nonumber \\&& \frac{V\mu_2}{X}(\frac{\mu_2}{Y}\big(e^{-rT}\{\frac{1}{T}\int_{0}^{T}(e^{tQ}(\rho_{23}\sigma_{2}(x)\sigma_{3}(x))dt\}\big) +  \frac{\mu_3}{Y}\big(e^{-rT}\{\frac{1}{T}\int_{0}^{T}(e^{tQ} \sigma_{3}^2(x)) dt\}\big)))\big(\frac{1}{\det R}\big(\frac{k Z}{Y}\big)\big) + \nonumber \\&& \big(\frac{Y^2}{x}(\frac{Y^2}{X}\big(e^{-rT}\{\frac{1}{T}\int_{0}^{T}(e^{tQ} \sigma_{1}^2(x)) dt\}\big) - \frac{V\mu_3}{X}\big(e^{-rT}\{\frac{1}{T}\int_{0}^{T}(e^{tQ}(\rho_{21}\sigma_{2}(x)\sigma_{1}(x))dt\}\big) + \nonumber \\&& \frac{V\mu_2}{X}\big(e^{-rT}\{\frac{1}{T}\int_{0}^{T}(e^{tQ}(\rho_{31}\sigma_{3}(x)\sigma_{1}(x))dt\}\big)) -  \frac{V\mu_3}{X}(\frac{Y^2}{X}\big(e^{-rT}\{\frac{1}{T}\int_{0}^{T}(e^{tQ}(\rho_{12}\sigma_{1}(x)\sigma_{2}(x))dt\}\big) - \nonumber \\&& \frac{V\mu_3}{X}\big(e^{-rT}\{\frac{1}{T}\int_{0}^{T}(e^{tQ} \sigma_{2}^2(x)) dt\}\big) +  \frac{V\mu_2}{X}\big(e^{-rT}\{\frac{1}{T}\int_{0}^{T}(e^{tQ}(\rho_{32}\sigma_{3}(x)\sigma_{2}(x))dt\}\big)) + \nonumber \\&& \frac{V\mu_2}{X}(\frac{Y^2}{X}\big(e^{-rT}\{\frac{1}{T}\int_{0}^{T}(e^{tQ}(\rho_{13}\sigma_{1}(x)\sigma_{3}(x))dt\}\big) -  \frac{V\mu_3}{X}\big(e^{-rT}\{\frac{1}{T}\int_{0}^{T}(e^{tQ}(\rho_{23}\sigma_{2}(x)\sigma_{3}(x))dt\}\big) + \nonumber \\&& \frac{V\mu_2}{X}\big(e^{-rT}\{\frac{1}{T}\int_{0}^{T}(e^{tQ} \sigma_{3}^2(x)) dt\}\big))\big)\big(\frac{1}{\det R}\big(-k\big(\frac{\mu_2 + \mu_3}{Y}\big) + Y\big)\big) + \big(\frac{V}{Z}(\frac{V}{Z}\big(e^{-rT}\{\frac{1}{T}\int_{0}^{T}(e^{tQ} \sigma_{1}^2(x)) dt\}\big) - \nonumber \\&& \frac{V\mu_3}{X}\big(e^{-rT}\{\frac{1}{T}\int_{0}^{T}(e^{tQ}(\rho_{21}\sigma_{2}(x)\sigma_{1}(x))dt\}\big) + \frac{V\mu_2}{X}\big(e^{-rT}\{\frac{1}{T}\int_{0}^{T}(e^{tQ}(\rho_{31}\sigma_{3}(x)\sigma_{1}(x))dt\}\big)) + \nonumber \\&& \frac{\mu_3}{Z}(\frac{V}{Z}\big(e^{-rT}\{\frac{1}{T}\int_{0}^{T}(e^{tQ}(\rho_{12}\sigma_{1}(x)\sigma_{2}(x))dt\}\big) + \frac{\mu_3}{Z}\big(e^{-rT}\{\frac{1}{T}\int_{0}^{T}(e^{tQ} \sigma_{2}^2(x)) dt\}\big) - \nonumber \\&& \frac{\mu_2}{Z}\big(e^{-rT}\{\frac{1}{T}\int_{0}^{T}(e^{tQ}(\rho_{32}\sigma_{3}(x)\sigma_{2}(x))dt\}\big))  \frac{\mu_2}{Z}(\frac{V}{Z}\big(e^{-rT}\{\frac{1}{T}\int_{0}^{T}(e^{tQ}(\rho_{13}\sigma_{1}(x)\sigma_{3}(x))dt\}\big) + - \nonumber \\&& \frac{\mu_3}{Z}\big(e^{-rT}\{\frac{1}{T}\int_{0}^{T}(e^{tQ}(\rho_{23}\sigma_{2}(x)\sigma_{3}(x))dt\}\big) -  \frac{\mu_2}{Z}\big(e^{-rT}\{\frac{1}{T}\int_{0}^{T}(e^{tQ} \sigma_{3}^2(x)) dt\}\big))\big) \nonumber \\&& \sqrt{1 - \frac{1}{(\det R)^2} \big(k^2 \frac{Z^2}{Y^2} + 
  \big(\big(-k\frac{\mu_2 + \mu_3}{Y}\big) + Y\big)^2\big)}\Bigg) 
   \times \big(\frac{1}{\det R}\big(-k\big(\frac{\mu_2 + \mu_3}{Y}\big) + Y\big)\big)
\end{eqnarray}

\begin{eqnarray}
  P (x_3) &=& \Bigg(\frac{V}{Z}((\frac{\mu_2}{Y}\big(e^{-rT}\{\frac{1}{T}\int_{0}^{T}(e^{tQ}(\rho_{21}\sigma_{2}(x)\sigma_{1}(x))dt\}\big) + \frac{\mu_3}{Y}\big(e^{-rT}\{\frac{1}{T}\int_{0}^{T}(e^{tQ}(\rho_{31}\sigma_{3}(x)\sigma_{1}(x))dt\}\big))+ \nonumber \\&&  \frac{\mu_3}{Z}(\frac{\mu_2}{Y}\big(e^{-rT}\{\frac{1}{T}\int_{0}^{T}(e^{tQ} \sigma_{2}^2(x)) dt\}\big) + \frac{\mu_3}{Y}\big(e^{-rT}\{\frac{1}{T}\int_{0}^{T}(e^{tQ}(\rho_{32}\sigma_{3}(x)\sigma_{2}(x))dt\}\big))  - \nonumber \\&& \frac{\mu_2}{X}(\frac{\mu_2}{Y}\big(e^{-rT}\{\frac{1}{T}\int_{0}^{T}(e^{tQ}(\rho_{23}\sigma_{2}(x)\sigma_{3}(x))dt\}\big) + \frac{\mu_3}{Y}\big(e^{-rT}\{\frac{1}{T}\int_{0}^{T}(e^{tQ} \sigma_{3}^2(x)) dt\}\big)))\big(\frac{1}{\det R}\big(\frac{k Z}{Y}\big)\big) + \nonumber \\&&
   (\frac{V}{Z}(\frac{Y^2}{X}\big(e^{-rT}\{\frac{1}{T}\int_{0}^{T}(e^{tQ} \sigma_{1}^2(x)) dt\}\big) - \frac{V\mu_3}{X}\big(e^{-rT}\{\frac{1}{T}\int_{0}^{T}(e^{tQ}(\rho_{21}\sigma_{2}(x)\sigma_{1}(x))dt\}\big) + \nonumber \\&& \frac{V\mu_2}{X}\big(e^{-rT}\{\frac{1}{T}\int_{0}^{T}(e^{tQ}(\rho_{31}\sigma_{3}(x)\sigma_{1}(x))dt\}\big)) +  \frac{\mu_3}{Z}(\frac{Y^2}{X}\big(e^{-rT}\{\frac{1}{T}\int_{0}^{T}(e^{tQ}(\rho_{12}\sigma_{1}(x)\sigma_{2}(x))dt\}\big) - \nonumber \\&& \frac{V\mu_3}{X}\big(e^{-rT}\{\frac{1}{T}\int_{0}^{T}(e^{tQ} \sigma_{2}^2(x)) dt\}\big) +  \frac{V\mu_2}{X}\big(e^{-rT}\{\frac{1}{T}\int_{0}^{T}(e^{tQ}(\rho_{32}\sigma_{3}(x)\sigma_{2}(x))dt\}\big)) - \nonumber \\&& \frac{\mu_2}{Z}(\frac{Y^2}{X}\big(e^{-rT}\{\frac{1}{T}\int_{0}^{T}(e^{tQ}(\rho_{13}\sigma_{1}(x)\sigma_{3}(x))dt\}\big) - \frac{V\mu_3}{X}\big(e^{-rT}\{\frac{1}{T}\int_{0}^{T}(e^{tQ}(\rho_{23}\sigma_{2}(x)\sigma_{3}(x))dt\}\big) + \nonumber \\&& \frac{V\mu_2}{X}\big(e^{-rT}\{\frac{1}{T}\int_{0}^{T}(e^{tQ} \sigma_{3}^2(x)) dt\}\big)))\big(\frac{1}{\det R}\big(-k\big(\frac{\mu_2 + \mu_3}{Y}\big) + Y\big)\big) +  (\frac{V}{Z}(\frac{V}{Z}\big(e^{-rT}\{\frac{1}{T}\int_{0}^{T}(e^{tQ} \sigma_{1}^2(x)) dt\}\big) + \nonumber \\&& \frac{\mu_3}{Z}\big(e^{-rT}\{\frac{1}{T}\int_{0}^{T}(e^{tQ}(\rho_{21}\sigma_{2}(x)\sigma_{1}(x))dt\}\big) - \frac{\mu_2}{Z}\big(e^{-rT}\{\frac{1}{T}\int_{0}^{T}(e^{tQ}(\rho_{31}\sigma_{3}(x)\sigma_{1}(x))dt\}\big)) + \nonumber \\&& \frac{\mu_3}{Z}(\frac{V}{Z}\big(e^{-rT}\{\frac{1}{T}\int_{0}^{T}(e^{tQ}(\rho_{12}\sigma_{1}(x)\sigma_{2}(x))dt\}\big) + \frac{\mu_3}{Z}\big(e^{-rT}\{\frac{1}{T}\int_{0}^{T}(e^{tQ} \sigma_{2}^2(x)) dt\}\big) - \nonumber \\&& \frac{\mu_2}{Z}\big(e^{-rT}\{\frac{1}{T}\int_{0}^{T}(e^{tQ}(\rho_{32}\sigma_{3}(x)\sigma_{2}(x))dt\}\big)) - \frac{\mu_2}{Z}(\frac{Y^2}{X}\big(e^{-rT}\{\frac{1}{T}\int_{0}^{T}(e^{tQ}(\rho_{13}\sigma_{1}(x)\sigma_{3}(x))dt\}\big) + \nonumber \\&& \frac{\mu_3}{Z}\big(e^{-rT}\{\frac{1}{T}\int_{0}^{T}(e^{tQ}(\rho_{23}\sigma_{2}(x)\sigma_{3}(x))dt\}\big) - \frac{\mu_2}{z}\big(e^{-rT}\{\frac{1}{T}\int_{0}^{T}(e^{tQ} \sigma_{3}^2(x)) dt\}\big)))\nonumber \\&& \sqrt{1 - \frac{1}{(\det R)^2} \big(k^2 \frac{Z^2}{Y^2} +
  \big(\big(-k\frac{\mu_2 + \mu_3}{Y}\big) +  Y\big)^2\big)}\Bigg)\times \sqrt{1 - \frac{1}{(\det R)^2} \big(k^2 \frac{Z^2}{Y^2} +
  \big(\big(-k\frac{\mu_2 + \mu_3}{Y}\big) +  Y\big)^2\big)} \nonumber
\end{eqnarray}
So our P(x) can be written as,
\begin{equation} 
    P (x) = P (x_1) + P (x_2) +P (x_3) - \big(e^{-rT} \times K_{strike price}\big) \nonumber
\end{equation}

\section{Numerical Example}

Stock of “CMS Energy Corporation” \citeauthor{Quantopian} (\citeyear{Quantopian}) is chosen as $S_{1}$, stock of “American Electric Power Company Inc” is chosen as $S_{2}$ and stock of “Entergy Corporation” is chosen as $S_{3}$. The data used are daily closing price of $S_{1}$, $S_{2}$ and $S_{3}$ in the time range 3rd May, 2018 till 2nd May, 2019.

To create our finite state Markov chain, we consider two states for each individual stock, defined as follows.
Considering stock $S_1$,
\[   
x_{1r} = \left\{
     \begin{array}{lll}
       \textbf{Up}, & \quad \text{when return} > \mu_1 & \text{128 observations}\\
       \textbf{Down}, &\quad \text{when return} \leq \mu_1 & \text{123 observations}\\
     \end{array}
\right.
\]

Similarly, for the stock $S_2$ we can divide the data points as,\newline
\[   
x_{2r} = \left\{
     \begin{array}{lll}
       \textbf{Up}, & \quad \text{when return} > \mu_2 & \text{138 observations}\\
       \textbf{Down}, &\quad \text{when return} \leq \mu_2 & \text{113 observations}\\
     \end{array}\right.
\]

Similarly, for the stock $S_3$ we can divide the data points as,\newline
\[   
x_{3r} = \left\{
     \begin{array}{lll}
       \textbf{Up}, & \quad \text{when return} > \mu_3 & \text{130 observations}\\
       \textbf{Down}, &\quad \text{when return} \leq \mu_3 & \text{121 observations}\\
     \end{array}\right.
\]

Now combining all the data points and considering that the combined state random variable takes three values, we get the following division:
\[   
x_{data} = \left\{
     \begin{array}{lll}
       \textbf{Up,} & \quad \text{when } \text{x}_{1} = \text{x}_{2} = \text{x}_{3} = \textbf{Up} & \text{93 observations}\\
       \textbf{Middle,} & \quad \text{otherwise} & \text{78 observations} \\
       \textbf{Down,} &\quad \text{when } \text{x}_{1} = \text{x}_{2} = \text{x}_{3} = \textbf{Down} & \text{80 observations} \\
     \end{array}\right.
\]

Using the functions from R studio we can calculate the transition probability matrix ($\Pi$) as,
\begin{center}
\begin{tabular}{ | m{1cm} | m{3cm}| m{3cm} | m{3cm}| } 
\hline
State & \textbf{Down} & \textbf{Middle} & \textbf{Up} \\
\hline
\textbf{Down} & 0.3250000 & 0.2875000 & 0.3875000 \\ 
\hline
\textbf{Middle} & 0.3636364 & 0.3376623 & 0.2987013 \\ 
\hline
\textbf{Up} & 0.2795699 & 0.3010753 & 0.4193548 \\ 
\hline
\end{tabular}
\end{center}
and the corresponding standard error of the probability matrix is given by
\begin{center}
\begin{tabular}{ | m{1cm} | m{3cm}| m{3cm} | m{3cm}| } 
\hline
State & \textbf{Down} & \textbf{Middle} & \textbf{Up} \\
\hline
\textbf{Down} & 0.06373774 & 0.05994789 & 0.06959705 \\ 
\hline
\textbf{Middle} & 0.06872081 & 0.06622103 & 0.06228353 \\ 
\hline
\textbf{Up} & 0.05482817 & 0.05689788 & 0.06715052 \\ 
\hline
\end{tabular}
\end{center}

The stationary probability matrix \textbf{p} can be calculated from the formula, 
\begin{equation}
    \textbf{p} \Pi = \textbf{p} \nonumber
\end{equation}
where 
\[\textbf{p} = 
\begin{bmatrix}
 \text{p}_D & \text{p}_M & \text{p}_U
\end{bmatrix}.
\]
Solving the equations on the variables we get,
\begin{center}
\begin{tabular}{| m{1cm} | m{2cm} | } 
\hline
$\text{p}_{D}$ & 0.32000\\
\hline
$\text{p}_{M}$ & 0.30783\\ 
\hline
$\text{p}_{U}$ & 0.37217\\ 
\hline
\end{tabular}
\end{center}

The mean returns of the stocks and the daily interest rate (assuming 10 $\%$ annual interest rate) is calculated as,

\begin{center}
\begin{tabular}{| m{1cm} | m{2cm} | } 
\hline
$mu_1$ & 0.000664\\
\hline
$mu_2$ & 0.000873\\ 
\hline
$mu_3$ & 0.0.000725\\ 
\hline
$r$ & 0.0004\\ 
\hline
\end{tabular}
\end{center}

We can evaluate the variance terms as,
\begin{equation}
    \text{P}_{\text{2var}}(i) = e^{-rT}\{\frac{1}{T}\int_{0}^{T}(\text{p}_{\text{iD}}\sigma^2_{2D} + \text{p}_{\text{iM}}\sigma^2_{2M} + \text{p}_{\text{iU}}\sigma^2_{2U})dt \} \nonumber
\end{equation}
where i = D,M,U is the initial state of the Markov chain. If we are uncertain about the initial state and we have only an initial probability distribution, say $(p_D,p_M,p_U)$ then the expression will be

\begin{equation}
  \text{P}_{\text{2var}} =   \text{p}_{\text{D}}\text{P}_{\text{2var}}(D) + \text{p}_{\text{M}}\text{P}_{\text{2var}}(M) + \text{p}_{\text{U}}\text{P}_{\text{2var}}(U) \nonumber
\end{equation}

Similarly other variance terms are
\begin{equation}
    \text{P}_{\text{1var}}(i) = e^{-rT}\{\frac{1}{T}\int_{0}^{T}(\text{p}_{\text{iD}}\sigma^2_{1D} + \text{p}_{\text{iM}}\sigma^2_{1M} + \text{p}_{\text{iU}}\sigma^2_{1U})dt \} \nonumber
\end{equation}

\begin{equation}
  \text{P}_{\text{1var}} =   \text{p}_{\text{D}}\text{P}_{\text{1var}}(D) + \text{p}_{\text{M}}\text{P}_{\text{1var}}(M) + \text{p}_{\text{U}}\text{P}_{\text{1var}}(U) \nonumber
\end{equation}

\begin{equation}
    \text{P}_{\text{3var}}(i) = e^{-rT}\{\frac{1}{T}\int_{0}^{T}(\text{p}_{\text{iD}}\sigma^2_{3D} + \text{p}_{\text{iM}}\sigma^2_{3M} + \text{p}_{\text{iU}}\sigma^2_{3U})dt \} \nonumber
\end{equation}

\begin{equation}
  \text{P}_{\text{3var}} =   \text{p}_{\text{D}}\text{P}_{\text{3var}}(D) + \text{p}_{\text{M}}\text{P}_{\text{3var}}(M) + \text{p}_{\text{U}}\text{P}_{\text{3var}}(U) \nonumber
\end{equation}

The values are shown in Table 1. 
\begin{table}
\caption{All the figures are in $10^{-6}$}\label{table1}
\begin{center}
\begin{tabular}{| m{1cm} | m{2cm} | } 
\hline
$\text{P}_{\text{1var}}$ & $42.978$\\
\hline
$\text{P}_{\text{2var}}$ & $43.275$\\ 
\hline
$\text{P}_{\text{3var}}$ & $40.240$\\ 
\hline
\end{tabular}
\end{center}
\end{table}

We can evaluate the covariance terms as,

\begin{equation}
   \text{P}_{\text{Cov(23)}}(i) = e^{-rT}\{\frac{1}{T}\int_{0}^{T}(\text{p}_{\text{iD}}\text{Cov}_{23D} + \text{p}_{\text{iM}}\text{Cov}_{23M} + \text{p}_{\text{iU}}\text{Cov}_{23U})dt \} \nonumber
\end{equation}
where i = D,M,U is the initial state of the Markov chain. If we are uncertain about the initial state and we have only a probability distribution, let say $(p_D,p_M,p_U)$  then the price is going be

\begin{equation}
  \text{P}_{\text{Cov(23)}} =   \text{p}_{\text{D}}\text{P}_{\text{Cov(23)}}(D) + \text{p}_{\text{M}}\text{P}_{\text{Cov(23)}}(M) + \text{p}_{\text{U}}\text{P}_{\text{Cov(23)}}(U) \nonumber
\end{equation}

Similarly other covariance terms are evaluated as
\begin{equation}
   \text{P}_{\text{Cov(31)}}(i) = e^{-rT}\{\frac{1}{T}\int_{0}^{T}(\text{p}_{\text{iD}}\text{Cov}_{31D} + \text{p}_{\text{iM}}\text{Cov}_{31M} + \text{p}_{\text{iU}}\text{Cov}_{31U})dt \} \nonumber
\end{equation}

\begin{equation}
  \text{P}_{\text{Cov(31)}} =   \text{p}_{\text{D}}\text{P}_{\text{Cov(31)}}(D) + \text{p}_{\text{M}}\text{P}_{\text{Cov(31)}}(M) + \text{p}_{\text{U}}\text{P}_{\text{Cov(31)}}(U) \nonumber
\end{equation}

and
\begin{equation}
   \text{P}_{\text{Cov(12)}}(i) = e^{-rT}\{\frac{1}{T}\int_{0}^{T}(\text{p}_{\text{iD}}\text{Cov}_{12D} + \text{p}_{\text{iM}}\text{Cov}_{12M} + \text{p}_{\text{iU}}\text{Cov}_{12U})dt \} \nonumber
\end{equation}

\begin{equation}
  \text{P}_{\text{Cov(12)}} =   \text{p}_{\text{D}}\text{P}_{\text{Cov(12)}}(D) + \text{p}_{\text{M}}\text{P}_{\text{Cov(12)}}(M) + \text{p}_{\text{U}}\text{P}_{\text{Cov(12)}}(U) \nonumber
\end{equation}

These values are shown in Table 2.
\begin{table}
\caption{All the figures are in $10^{-6}$}\label{table2}
\begin{center}
\begin{tabular}{| m{3cm} | m{2cm} | } 
\hline
$\text{P}_{\text{Cov(23)}} =  \text{P}_{\text{Cov(32)}}$ & $41.234$\\
\hline
$\text{P}_{\text{Cov(31)}} = \text{P}_{\text{Cov(13)}}$ & $39.477$\\ 
\hline
$\text{P}_{\text{Cov(12)}} = \text{P}_{\text{Cov(21)}}$ & $40.911$\\ 
\hline
\end{tabular}
\end{center}
\end{table}

\subsection{Numerical example for swap given by trace}

As mentioned in the introduction, our first candidate measure of generalized variance is the trace of the covariance matrix which is nothing but the sum of the individual variances. Intuitively, this is the variance of the return of a portfolio comprising one unit of each of the stocks, assuming them to be uncorrelated.
We have already calculated the expected value of the variances in Table \ref{table1}. So we can calculate the price of the swap as,

\begin{equation} 
    P_{trace}(x) = \big(\text{P}_{\text{1var}} + \text{P}_{\text{2var}} + \text{P}_{\text{3var}}\big) - e^{-rT} K_{\text{strike price}} \nonumber
\end{equation}

\begin{equation} 
    P_{trace}(x) = 42.978 + 43.275 + 40.240 - e^{-rT} K_{\text{strike price}} \nonumber
\end{equation}

\begin{equation} 
    P_{trace}(x) = 126.493 - e^{-rT} K_{\text{strike price}} \nonumber
\end{equation}
The swap is written in terms of 1 million units of the trace. Considering the strike price as 90 and duration is for 3 months i.e. T = 63 we can finally calculate the trace swap as,
\begin{equation} 
    P_{\text{trace}} = 126.493
- \big(e^{-(0.0004)\times63} \times 90\big) = 126.493 - 87.760 = 38.733 \nonumber
\end{equation}
\begin{equation} 
    P_{\text{trace}} = 38.733 \nonumber
\end{equation}

\subsection{Numerical example for swap given by largest eigenvalue}

The second candidate measure of generalised variance considered here is the maximum eigenvalue which is the magnitude of the biggest component of the orthogonalised system for the return covariance matrix. As the return distributions are correlated (the covariance matrix is not diagonal), this biggest component will be significantly larger than the individual variances. This is considered as we are interested in managing the variance, so swapping for the biggest component is a safe strategy to adopt.

Considering T = 63 we do the following calculations,
\[
\textbf{A}=
\begin{bmatrix}
    0.000664 & 1 \\
    0.000873 & 1\\
    0.000725 & 1 \\
\end{bmatrix}
\]
Doing a QR decomposition of A, we get
\[
\textbf{P}_1=
\begin{bmatrix}
    0.505023 & 0.6551103 \\
    0.6639542 & -0.7108661\\
    0.5514677 & 0.2559293
\end{bmatrix}
\]
\[
\textbf{R}=
\begin{bmatrix}
    0.001315 & 1.720445 \\
    0 & 0.2001735\\
\end{bmatrix}
\]
\[
\textbf{P}_2=
\begin{bmatrix}
    0.561945  \\
    0.232022 \\
    -0.793967 
\end{bmatrix}
\]
Therefore P is,
\[
\textbf{P}=
\begin{bmatrix}
    0.505023 & 0.6551103 & 0.561945\\
    0.6639542 & -0.7108661 & 0.232022\\
    0.5514677 & 0.2559293 & -0.793967
\end{bmatrix}
\]
\[
\textbf{R}^{-T}=
\begin{bmatrix}
    760.4563 & 0 \\
    -6536.689733 & 4.995656\\
\end{bmatrix}
\]
Given that,

\[ \bf{b} = 
\begin{bmatrix}
   0.0007 \\
    1 
\end{bmatrix}
\]

\[
\textbf{q} = 
\begin{bmatrix}
     0.5322577 \\
  0.4210340
\end{bmatrix}
\]
\[
\textbf{r} = 
\begin{bmatrix}
     0.7344604
\end{bmatrix}
\]
Therefore we can calculate \textbf{w(t)}
\[
\textbf{w(t)} = 
\begin{bmatrix}
       0.9569597 \\
  0.2239916 \\  -0.1822877
\end{bmatrix}
\]

Note that, since the expected returns of the portfolio are quite spread out, in the optimal solution we are getting short position for one stock which has a higher expected return and long positions for the other two stocks.

\[ \big(e^{-rT}\{\frac{1}{T}\int_{0}^{T}(e^{tQ}(\Omega)dt\}\big) = 
\begin{bmatrix}
    \text{P}_{\text{1var}} & \text{P}_{\text{Cov(12)}} & \text{P}_{\text{Cov(13)}}\\
    \text{P}_{\text{Cov(21)}} & \text{P}_{\text{2var}} & \text{P}_{\text{Cov(23)}}\\
    \text{P}_{\text{Cov(31)}} & \text{P}_{\text{Cov(32)}} & \text{P}_{\text{3var}}
\end{bmatrix}
\]
Thus the price of the eigenvalue swap is given by,
\begin{equation} 
    P_{\text{eigenvalue}} = \begin{bmatrix}
       0.9569597 \\
  0.2239916 \\  -0.1822877
\end{bmatrix}^{T}
    \begin{bmatrix}
    \text{P}_{\text{1var}} & \text{P}_{\text{Cov(12)}} & \text{P}_{\text{Cov(13)}}\\
    \text{P}_{\text{Cov(21)}} & \text{P}_{\text{2var}} & \text{P}_{\text{Cov(23)}}\\
    \text{P}_{\text{Cov(31)}} & \text{P}_{\text{Cov(32)}} & \text{P}_{\text{3var}}
\end{bmatrix}
\begin{bmatrix}
       0.9569597 \\
  0.2239916 \\  -0.1822877
\end{bmatrix}
- \big(e^{-rT} \times K_{strike price}\big) \nonumber
\end{equation}
Putting the values we get,

\begin{equation} 
    P_{\text{eigenvalue}} = \begin{bmatrix}
       0.9569597 \\
  0.2239916 \\  -0.1822877
\end{bmatrix}^{T}
    \begin{bmatrix}
    42.978 & 40.911 & 39.477\\
    40.911 & 43.275 & 41.234\\
    39.477 & 41.234 & 40.240
\end{bmatrix}
\begin{bmatrix}
       0.9569597 \\
  0.2239916 \\  -0.1822877
\end{bmatrix}
- \big(e^{-rT} \times K_{strike price}\big) \nonumber
\end{equation}
\begin{equation} 
     = \begin{bmatrix}
       43.095 & 41.327 & 39.678
\end{bmatrix}
\begin{bmatrix}
       0.9569597 \\
  0.2239916 \\  -0.1822877
\end{bmatrix}
- \big(e^{-rT} \times K_{strike price}\big) \nonumber
\end{equation}
\begin{equation} 
    P_{\text{eigenvalue}} = 43.264
- \big(e^{-rT} \times K_{strike price}\big) \nonumber
\end{equation}
The swap is written in terms of 1 million units of the eigenvalue. Considering the strike price as 30 we can finally calculate the eigenvalue swap as,
\begin{equation} 
    P_{\text{eigenvalue}} = 43.264
- \big(e^{-(0.0004)\times63} \times 30\big) = 43.264 - 29.253 = 14.011 \nonumber
\end{equation}
\begin{equation} 
    P_{\text{eigenvalue}} = 14.011 \nonumber
\end{equation}

\section{Conclusion}

In this paper we have presented a new approach for pricing swaps defined on two important measures of generalized variance,  namely the maximum eigenvalue and trace of the covariance matrix of the returns on assets involved. The objective is to price generalized variance swaps for financial markets with Markov-modulated volatilities. We have considered multiple assets in the portfolio for  theoretical purpose and demonstrated the theoretical approach with the help of numerical examples taking three stocks in the portfolio. The results derived in this paper are the comparison between the swaps defined by the trace and the eigenvalue. In the numerical examples of swaps priced, the price of the trace swap is more than that of the value for the eigenvalue swap. This can be justified as in the maximum eigenvalue swap it is not only the variance of the stocks which is responsible for the price determination, it is also the covariance terms and the expected value of the stocks' return which are present in the price of the maximum eigenvalue swap which makes the price for the maximum eigenvalue swap less. So, we can say that for the same stocks we should prefer the maximum eigenvalue swap as compared to trace swap as the price of the swap is less. Moreover, the results obtained in this paper have important implications for their use in the commodity sector as volatility in the commodity markets, agricultural in particular, are often related through natural causes. This would be an important area where such swaps would be useful for hedging risk. In our future work we also aim to incorporate an often observed phenomenon in the returns, namely jumps \citeauthor{broadie2} (\citeyear{broadie2}). This would render the usual Ito formulation unsatisfactory. One can apply the well known Levy process to the returns in such a case \citeauthor{sengupta} (\citeyear{sengupta}). It would be an important extension to define and price swaps on measures of generalized variance in this scenario.



\begin{thebibliography}{}
%
%
\bibitem[Benth {\it et al.}(2007)]{benth} F. E. Benth, M. Groth \& R. Kufakunesu (September 2007) {\it Valuing Volatility and Variance Swaps for a Non-Gaussian Ornstein-Uhlenbeck Stochastic Volatility Model}. Applied Mathematical Finance, Vol. 14, No. 4, 347-363.

\bibitem[Bossu(2005)]{bossu1} S. Bossu (2005) {\it Arbitrage Pricing of Equity Correlation Swaps}. JPMorgan Equity Derivatives, Working paper.

\bibitem[Bossu(2007)]{bossu2} S. Bossu (2007) {\it A New Approach For Modelling and Pricing Correlation Swaps. Equity
Structuring - ECD London}, Working paper.

\bibitem[Broadie and Jain(2008a)]{broadie1} M. Broadie \& A. Jain (2008a) {\it Pricing and Hedging Volatility Derivatives}. The Journal of Derivatives, Vol. 15, No. 3, pp. 7-24.

\bibitem[Broadie and Jain(2008b)]{broadie2} M. Broadie \& A. Jain (2008b) {\it The Effect of Jumps and Discrete Sampling on Volatility and Variance Swaps}. International Journal of Theoretical and Applied Finance, Vol.11, No.8. pp. 761-797.

\bibitem[Carr {\it et al.} (2005)]{carr1} P. Carr, H. Geman, D. B. Madan \& M. Yor  (2005) {\it Pricing options on realized variance}. Finance Stochast. 9, 453-475.

\bibitem[Carr and Lee(2007)]{carr2} P. Carr \& R. Lee (2007) {\it Realized volatility and variance: Options via swaps}. Bloomberg LP and University of Chicago. Available at:
http://math.uchicago.edu/˜rl/OVSwithAppendices.pdf.

\bibitem[Elliott and  Swishchuk(2007)]{swischuk1} R. Elliott \& A. V. Swishchuk  (2007) {\it Pricing Options and Variance Swaps in Markov Modulated Brownian Markets. In: ‘Hidden Markov Model in Finance’}, Eds. R. Mamon and R. Elliott, Springer.

\bibitem[Fonseca  {\it et al.} (2009)]{fonseca} J. Da Fonseca, F. Ielpo \& M. Grasselli (2009) {\it Hedging (Co)Variance Risk with Variance Swaps}. Available at SSRN: http://ssrn.com/abstract=1341811.

\bibitem[Gander {\it et al.}(1991)]{Ref3}
Walter Gander, Gene H. Golub \& Urs von Matt  (1991) {\it A Constrained Eigenvalue Problem, Numerical Linear Algebra}, Digital Signal Processing and Parallel Algorithms, Vol 70, 677-686.

\bibitem[Habtemicael and SenGupta (2016)]{sengupta}
Semere Habtemicael \& Indranil SenGupta  (2016) {\it Pricing covariance swaps for Barndorff-Nielsen and Shephard process driven financial markets},  Annals of Financial Economics, Vol 11, pp. 1650012.

\bibitem[Salvi and Swishchuk(2012)]{swischuk2} G. Salvi \& A. V. Swishchuk (2012) {\it Pricing of Variance, Volatility, Covariance and Correlation Swaps in a Markov-modulated Volatility Model}. Preprint.

\bibitem[Quantopian(2018)]{Quantopian} {\it https://www.quantopian.com}, Accessed: 2018-10-10.






\end{thebibliography}


\end{document}